\documentclass[aps,pra,twocolumn,superscriptaddress,showpacs,preprintnumbers,amsmath,amssymb]{revtex4-1}

\usepackage{graphicx}
\usepackage{dcolumn}
\usepackage{bm}
\usepackage{enumerate}
\usepackage{mathrsfs}
\usepackage{amsmath}
\usepackage{amsfonts}

\usepackage{subfigure}
\usepackage{tabularx}

\usepackage{epstopdf}
\usepackage[titletoc]{appendix}
\usepackage[colorlinks=true,breaklinks=true,linkcolor=blue,citecolor=blue,urlcolor=blue]{hyperref}
\usepackage{color}
\usepackage{array,mathtools,amssymb,booktabs}

\usepackage{multirow}

\newcolumntype{C}{>{$}c<{$}}
\AtBeginDocument{
	\heavyrulewidth=.08em
	\lightrulewidth=.05em
	\cmidrulewidth=.03em
	\belowrulesep=.65ex
	\belowbottomsep=0pt
	\aboverulesep=.4ex
	\abovetopsep=0pt
	\cmidrulesep=\doublerulesep
	\cmidrulekern=.5em
	\defaultaddspace=.5em
}

\begin{document}
	
\bibliographystyle{apsrev4-2} 
	
\def\rtx@apspra{\class@info{APS journal PRA selected}}
	
\title{Frequency super-resolution with quantum environment engineering \\in a weakly coupled nuclear-spin system}	
	
\author{Tianzi Wang}
\affiliation{School of Physics and Technology, Wuhan University, Wuhan, Hubei 430072, China}
\affiliation{Hefei National Laboratory, Hefei, Anhui 230088, China}
\author{Qian Cao}
\affiliation{School of Physics and Technology, Wuhan University, Wuhan, Hubei 430072, China}
\author{Peng Du}
\affiliation{School of Physics and Technology, Wuhan University, Wuhan, Hubei 430072, China}
	
\author{Wenxian Zhang}
\email[Contact author: ]{wxzhang@hznu.edu.cn}
\affiliation{School of Physics, Hangzhou Normal University, Hangzhou, Zhejiang 311121, China}
\affiliation{School of Physics and Technology, Wuhan University, Wuhan, Hubei 430072, China}
	
\date{\today}
	
\begin{abstract}
	Optical super-resolution has been widely employed to beat spatial diffraction limit, which is often stated by Abbe-Rayleigh criterion. Analogously, we propose a frequency super-resolution method, which beats conventional spectral resolution limit often approximated by full width half maximum of the spectral peak, $\Gamma$. This method utilizes recently developed quantum environment engineering technique. With numerical simulations and experiments, we demonstrate the frequency super-resolution method in a three-nuclear-spin system (Trifluoroiodoethylene), by successfully decomposing a thermal state spectrum of the spin $F_3$ into four peaks of engineered pseudo-pure states of the quantum environment. The ultimate frequency resolution reaches $\sim 0.005 \,\Gamma$. This method is potentially useful in spectral decomposition of weakly coupled nuclear spin systems and might be improved further to acquire finer frequency super-resolution by employing more advanced quantum techniques.
\end{abstract}
	
\maketitle
	
\section{Introduction}
	
Optical spectrum (emission or absorption) is an efficient and powerful tool to detect microscopic world, such as atomic and molecular structure by infrared and visible light~\cite{B_C_Fawcett_1981, RevModPhys.68.1015}, chemical and biological structure by microwave and radio frequency~\cite{slichter1996principles}. Fine and hyperfine structures are pursued by pushing very extreme laboratory conditions, e.g., lowing the temperature to nanokelvin~\cite{ORRIT1992141}, raising the vacuum to ultrahigh level (nanobar)~\cite{10.1063/1.2794227}, and increasing magnetic field to tens of Tesla~\cite{tentesla, Yanagisawa_2022}. However, the spectrum is ultimately limited by a ``diffraction limit", i.e., two adjacent emission (or absorption) lines are not discernible if their separation is smaller than their line width $\Gamma$ which is due to homogeneous and inhomogeneous line broadening~\cite{doi:10.1126/science.257.5067.189, PhysRevX.6.031033}. To improve the spectral resolution, numerical methods rooted in information science are employed~\cite{MUSIC, DMFIT}. For instance, the multiple signal classification technique is widely used to precisely determine the peak positions of many-peak signals~\cite{highresMUSIC}. However, these numerical multiple peak fitting methods rely heavily on the assumptions of peak number and peak profile which are often conjectured without direct and physical justification.
 
An alternative way to beat the frequency resolution limit may be borrowed from the idea of optical super-resolution~\cite{HUSZKA20197, Hell_2015, rust2006sub,Bo2008, palmBetzig}. In conventional optics, two point light sources are not spatially resolvable on the image plane if their distance is smaller than the Abbe-Rayleigh criterion (spatial diffraction limit)~\cite{doi:10.1142/5833, Jiang2019}. A clever idea in photoactivated localization microscopy (PALM) by switching on and off local fluorescence is employed to beat the diffraction limit and reaches $\sim$10 nm spatial resolution with the light wave length $\sim$500 nm~\cite{palmBetzig}. In fact, such an optical super-resolution is realized by introducing an extra dimension of time where the multiple peaks are stochastically separated but keep the Airy disc unchanged for each point light source without considering numerical method to reduce the Airy disc. In analogy, we may purposefully introduce an extra dimension to break the spectral resolution limit of an overlapped spectrum and realize frequency super-resolution (FRESURE) , by ``switching on" a single peak and ``turning off" the rest peaks but keeping the line width untouched (find more details in Appendix~\ref{apd:diffraction}).

Our main idea is to realize FRESURE by introducing the extra dimension, quantum environment (QE) state. It is known that a quantum system always couples with its environment, which may be divided into quantum and classical part. The interaction of the system with its QE often introduces line splitting, e.g., hyperfine splitting by electron and nuclear spin coupling, while the interaction with its classical environment causes line broadening (line width) in the nuclear spin system we consider here~\cite{PhysRevA.74.040502}. Once the line width is larger than the line splitting (under frequency resolution limit), one would fail to resolve the splitting. With the advancement of quantum technology for recent decades, it now becomes possible to precisely manipulate and engineer a certain QE state~\cite{dnpreview, Ryan2005, Bhole_2020, cao2023control}, which exhibits a single peak with the same line width (similar to PALM but in a definite way). Therefore, FRESURE may be realized by introducing a set of engineered QE states, serving as the additional dimension.

In this paper we put forward a theory of FRESURE for a weakly coupled three-nuclear-spin system, where a spin is considered as the quantum system (emitting radio frequency light) and the other two spins the QE~\cite{Majeed2019}. Numerical simulations and experimental results of nuclear magnetic resonance (NMR) further confirm the validity of the theory. Certainly, super-resolved spectrum will find many scientific and practical applications in much broader areas, not only in physics but also in chemistry, biology, medicine, and magnetic resonance imaging.

The paper is organized as follows. We introduce in Sec.~\ref{sec:derive} the Hamiltonian of the three-nuclear-spin system and derive the theory of physical spectral decomposition and FRESURE. In Sec.~\ref{sec:numerical}, we numerically confirm the validity of theoretic predictions. With a NMR simulator  of SpinQ Triangulum~\cite{ZengBei}, experimental results are shown in Sec.~\ref{sec:experiment}, which agree well with theoretical results. Finally, in Sec.~\ref{sec:conc} we summarize and discuss further experiments in the future.
	
	
\section{Physical spectral decomposition and FRESURE with QE engineering}
\label{sec:derive}
We consider a typical spin system with three weakly coupled spin-1/2 nuclei $F_i$ ($i=1,2,3$)~\cite{Heidebrecht2006}. Its Hamiltonian in a strong bias magnetic field $B_0$ along $z$ axis can be in principle separated into three parts (we set $\hbar=1$)
\begin{equation}
	\begin{aligned}
		H &= H_S + H_Q + H_{SQ},\\
		H_S &= {1\over 2} (\omega_0+\delta_3+\eta)\sigma_{3z}, \\
		H_Q &= {1\over 2}\sum_{i=1}^2(\omega_0+\delta_i+\eta)\sigma_{iz} + {J_{12}\over 4} {\bf \sigma_1}\cdot {\bf \sigma_2}, \\
		H_{SQ} &= {1\over 4}\sum_{i=1}^2J_{i3}{\bf \sigma_i}\cdot {\bf \sigma_3},
	\end{aligned}
	\label{eq:orginalH}
\end{equation}
where $H_S, H_Q$ and $H_{SQ}$ are respectively the system Hamiltonian, the QE Hamiltonian, and the system-environment coupling. The Larmor frequency of the spins is $\omega_0=\gamma B_0$ with $\gamma$ the nuclear gyromagnetic ratio, $\delta_i$ is the chemical shift of the $i$th nuclear spin, $J_{ij}$ characterizes the spin-spin coupling constant, and $\eta=\gamma  b_z$ denotes the noise from inhomogeneous stray magnetic field $\bf b$. Typical NMR experiment satisfies $\omega_0 \gg |\delta_i|, |J_{ij}|, |\bf b|$ thus the effects of $b_x$ and $b_y$ are averaged out and neglected. In addition, we assume weak coupling, i.e., $|\delta_i-\delta_j|\gg |J_{ij}|$, so that the Hamiltonian is further simplified to the secular coupling form after neglecting the transversal coupling terms wth $\sigma_{ix}\sigma_{jx}$ and $\sigma_{iy}\sigma_{jy}$
\begin{eqnarray}
	H_E\approx {1\over 2}\sum_{i=1}^3(\omega_0+\delta_i+\eta)\sigma_{iz}+{1\over 4}\sum_{i<j}J_{ij} \sigma_{iz}\sigma_{jz}.
	\label{eq:scalarH}
\end{eqnarray}
	
The thermal equilibrium state of the spin system is $\rho_T=e^{-H/(k_BT)}/{\text{Tr}}(e^{-H/(k_BT)})$. In the high temperature limit $k_B T \gg \omega_0 \gg |\delta_i|, |J_{ij}|$ for typical NMR spin systems, the thermal equilibrium state becomes
$
\rho_T = (I/8)+(p/2)\sum_{i=1}^3\sigma_{iz}
$
where $p \approx -\omega_0/(8k_BT)$ and $I$ is the $8 \times 8$ identity matrix.
A thermal initial state is prepared by performing a $\pi/2$ rotation of spin $F_3$ along $y$ axis 
\begin{eqnarray}
	\rho_T(0) = \frac{1}{8}I+\frac{p}{2}(\sigma_{1z}+\sigma_{2z}+\sigma_{3x}).
	\label{eq:ini thm}
\end{eqnarray}
	
Starting from the thermal equilibrium state, a pseudo pure state (PPS) is prepared with time averaging method~\cite{NI1997pps, PhysRevA.94.012312DJF, ZengBei},
$\rho_{PPS}=(1-q)I/8+q|000\rangle\langle000|$. Since the identity matrix $I$ has no contribution to the free induction decay (FID) signal, the PPS acts like a pure state~\cite{CORY199882}. Four initial PPSs may be prepared by flipping the spin $F_1$ or $F_2$ with $\pi$ pulses and rotating spin $F_3$ with a $\pi/2$ pulse, 
\begin{equation}
	\begin{aligned}
		\rho_A(0)&=\frac{1-q}{8}I+q|00\rangle\langle00|\otimes(I_2+\sigma_{3x})/2, \\
		\rho_B(0)&=\frac{1-q}{8}I+q|01\rangle\langle01|\otimes(I_2+\sigma_{3x})/2, \\
		\rho_C(0)&=\frac{1-q}{8}I+q|10\rangle\langle10|\otimes(I_2+\sigma_{3x})/2, \\
		\rho_D(0)&=\frac{1-q}{8}I+q|11\rangle\langle11|\otimes(I_2+\sigma_{3x})/2.
	\end{aligned}
	\label{eq:PPSABCD}	
\end{equation} 
where $I_2$ is a $2 \times 2$ identity matrix of spin $F_3$.
	
Starting from an initial spin state $\rho(0)$, the system evolves under the effective Hamiltonian $H_E$, Eq.~(\ref{eq:scalarH}). At time $t$, the FID signal of spin $F_3$ is $s(t)\equiv \langle\sigma_{3x}\rangle=\text{Tr}(U_E\rho(0)U_E^\dagger\sigma_{3x})$, where $U_E=e^{-itH_E}$. It is straightforward to calculate the FID signal for the four initial PPSs 
\begin{equation}
	\begin{aligned}
		s_A(t)=&\text{Tr}(U_E\rho_A(0)U_E^\dagger\sigma_{3x})=q\cos(\omega_A t), \\
		s_B(t)=&\text{Tr}(U_E\rho_B(0)U_E^\dagger\sigma_{3x})=q\cos(\omega_B t), \\
		s_C(t)=&\text{Tr}(U_E\rho_C(0)U_E^\dagger\sigma_{3x})=q\cos(\omega_C t), \\
		s_D(t)=&\text{Tr}(U_E\rho_D(0)U_E^\dagger\sigma_{3x})=q\cos(\omega_D t).
	\end{aligned}
	\label{eq:PPS sig}
\end{equation}
where $\omega_A = \alpha+\beta+\gamma, \omega_B = \alpha+\beta-\gamma, \omega_C = \alpha-\beta+\gamma$ and $\omega_D = \alpha-\beta-\gamma$ with $\alpha=\omega_0+\delta_3+\eta$, $\beta=J_{13}/2$, and $\gamma=J_{23}/2$. Obviously, the FID signal starting from each of four initial PPSs exhibits different and single oscillation frequency. If the initial state is the thermal initial state, the FID is 
$s_T(t)=\text{Tr}(U_E\rho_T(0)U_E^\dagger\sigma_{3x})
		=p[\cos(\omega_A t)+\cos(\omega_B t)+\cos(\omega_C t)+\cos(\omega_D t)]$. 
It's obvious that $s_T(t)$ consists of four oscillations, each of which represents a peak after Fourier transform. Interestingly, one immediately finds 
\begin{equation}
	s_T(t)=\frac{p}{q}(s_A(t)+s_B(t)+s_C(t)+s_D(t)).
	\label{eq:sum}
\end{equation}
Such a relation indicates that we can decompose the spectrum from the thermal initial state into the spectra from four different initial PPSs which are actually prepared by  QE engineering on spins $F_1$ and $F_2$.
	
Different from usual numerical multiple-peak-fitting methods~\cite{DMFIT}, such a spectral decomposition is physical and independent of the distribution of magnetic noise $\eta$, i.e., the profile of the peaks. It is especially useful if the line widths of adjacent peaks are larger than the peak splitting, which is nothing but FRESURE (similar to optical super-resolution). Of course, the weak coupling condition $|\delta_i-\delta_j| \gg |J_{ij}|$ is required in principle in order for the physical spectral decomposition to be valid.

\section{FRESURE with numerical simulations}
\label{sec:numerical}
\begin{figure}[tb]
	\centering
	\subfigure[]{\includegraphics[width=0.4\linewidth]{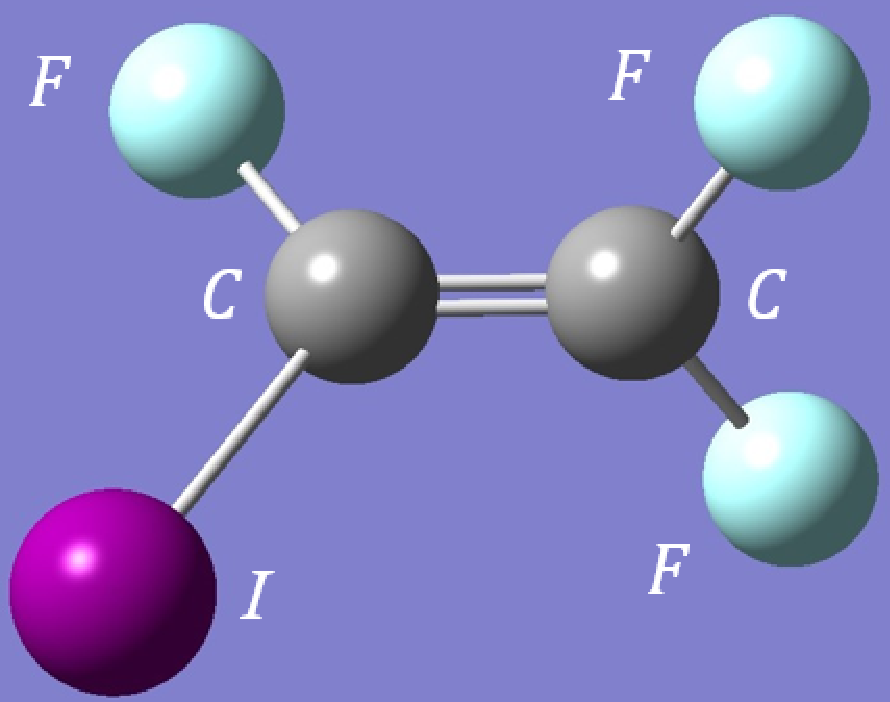}\label{fig:C2F3I}}
	\subfigure[]{\includegraphics[width=0.5\linewidth,height=0.11\textheight]{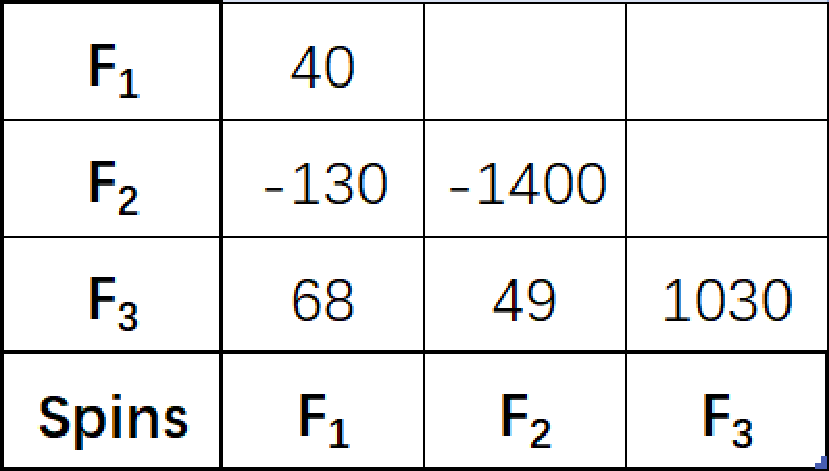}\label{fig:table}}

	\caption{(a) Structure of molecule $C_2F_3I$. The nuclear spins of three fluorine atoms are spin-1/2. (b) Chemical shifts (diagonal) and coupling constants (off-diagonal) for the fluorine spins, in the unit of Hz. We set the second peak of spin $F_1$ as $0$ Hz (see Appendix~\ref{apd:fid} for details).}
	\label{fig:structure}
\end{figure}

Under the weak coupling assumption, we have simplified the original Heisenberg spin coupling to the Ising coupling, as shown in Eq.~(\ref{eq:scalarH}). However, the ratio between the coupling constant and the chemical shift's difference $|J_{ij}/(\delta_i-\delta_j)| \sim 0.1$ is not small enough in the case of Trifluoroiodoethylene, $C_2F_3I$, as shown in Fig.~\ref{fig:structure}. Therefore, we employ numerical simulations to check the validity of the physical spectral decomposition. In our numerics, we adopt the Hamiltonian Eq.~(\ref{eq:orginalH}) with the Zeeman splitting of the strong bias magnetic field $\omega_0 = 39.638$ MHz. We assume $\eta$ follows a probability distribution with a Lorentzian profile 
$
f(\eta) =  (\Gamma/2\pi) /[\eta^2 + (\Gamma/2)^2],
\label{eq:lorentz}
$
where $\Gamma = 40$ Hz. Other parameters we use are given in Fig.~\ref{fig:table}.

\begin{figure}[tb]
	\centering
	\includegraphics[width=3.25in]{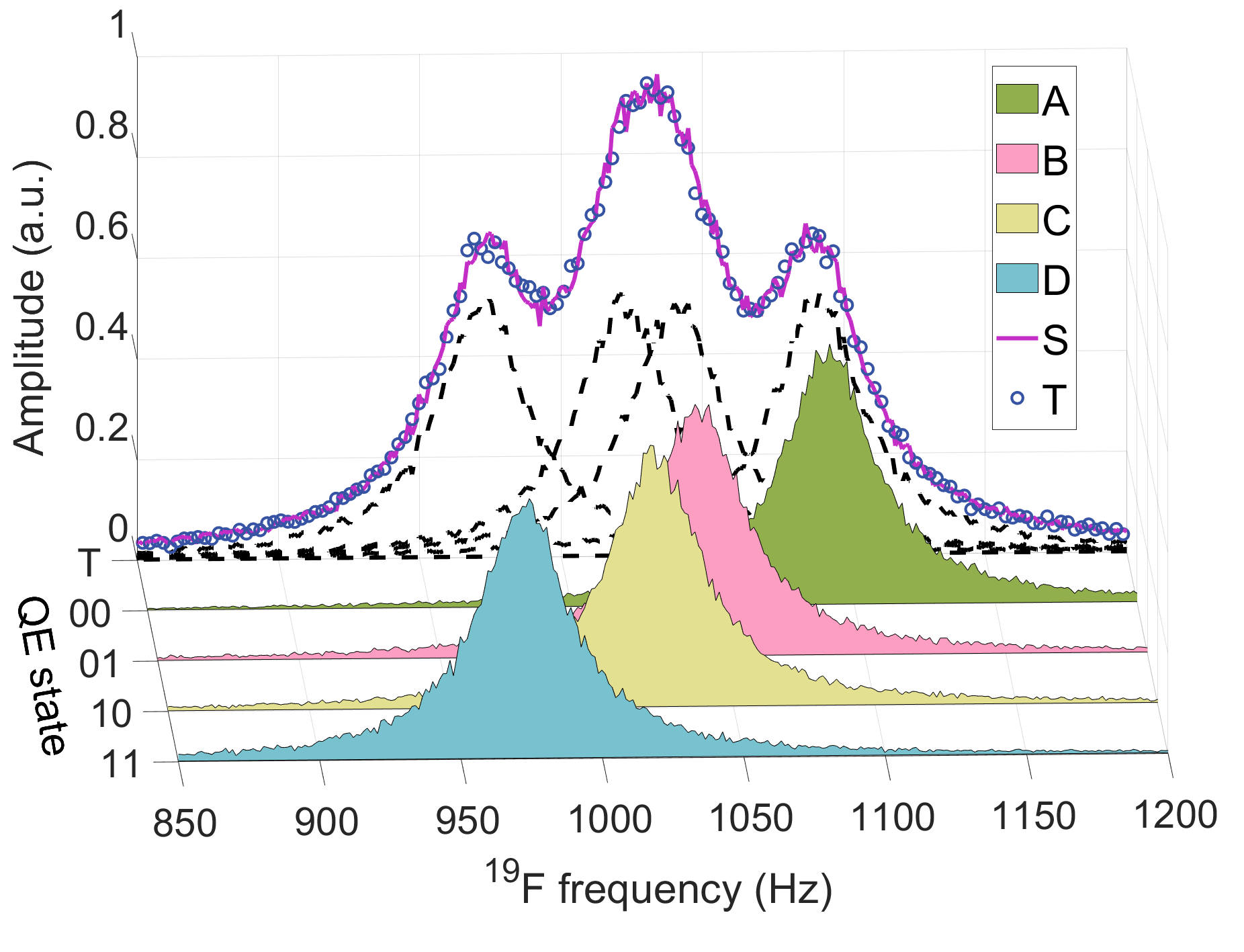}
	\caption{Numerical results of physical spectral decomposition. Four dashed lines denote the spectra obtained from FID signal $s_A, s_B, s_C, s_D$ of initial PPSs. The sum (purple solid line) of four dashed lines agrees well with the spectrum from the thermal initial state $s_T$ (blue circles). The stray magnetic-field noise $\eta$ follows a Lorentzian distribution with a full width at half maximum (FWHM) $\Gamma = 40$ Hz. Four PPS's spectra are also plotted in an additional dimension denoted by the QE state.}
	\label{fig:numerical}
\end{figure}

With these specific parameters and a given initial density matrix of three spins, we calculate the FID signal at different evolution time $t$ under the Hamiltonian Eq.~(\ref{eq:orginalH}). The FID signal is actually averaged over $10^5$ runs of random $\eta$ which obeys the Lorentzian probability distribution function. A spectrum is then obtained by performing a fast Fourier transform (FFT) of the averaged FID signal. We plot in Fig.~\ref{fig:numerical} the spectra with five initial states, the thermal initial state and four initial PPSs (A, B, C, and D). As a comparison, the sum spectrum of four initial PPSs is also presented. Due to their independence of each other, four PPS's spectra may be also plotted in an additional dimension denoted by the QE state. 
	
From Fig.~\ref{fig:numerical}, we see that the sum spectrum almost coincides with the thermal spectrum, indicating that the thermal spectrum can be decomposed into four PPS spectra with equal weight in the weak coupling regime we consider here. Moreover, the peak positions of the four PPSs' spectra ($1091.3, 1040.0, 1020.5, 974.1$) agree well with the analytical results ($1088.5, 1039.5, 1020.5, 971.5$) Eq.~(\ref{eq:PPS sig}), manifesting the fact that small deviation between the original Heisenberg coupling and the effective Ising coupling is negligible.
	
In contrast to the numerical fitting method ``dmfit'' (see Fig.~\ref{fig:dmfitnum34} in  Appendix~\ref{apd:dmfit} for fitting results), the decomposition into four PPSs is purely physical (shown in Fig.~\ref{fig:numerical})  and there is no need to carefully adjust any parameter. More importantly, by engineering the QE state as an additional dimension, we are able to definitely distinguish the close peaks B and C whose distance is smaller than their width $\Gamma$. In the spirit of optical super-resolution~\cite{Hell_2015, Henriques2011, nl20}, we thus realize FRESURE by QE engineering.

	
\section{FRESURE in experiments}
\label{sec:experiment}
To check the above analytical and numerical results on the spectral decomposition and the realization of FRESURE, we carry out NMR experiments with a real physical system composed of three $^{19}$F nuclear spins of Trifluoroiodoethylene ($C_2F_3I$). The structure of the molecule is shown in Fig.~\ref{fig:C2F3I}. The natural Hamiltonian of the system is written as Eq.~(\ref{eq:orginalH}) with parameters given in Fig.~\ref{fig:table}~\cite{Dujiangfeng2016}. The experiments are performed at a fixed temperature 35$^{\circ}$C, using a SpinQ Triangulum ($B_0= 0.97$ Tesla)~\cite{ZengBei}.
	
The experiments are divided into three steps in general: thermal equilibrium state preparation after long enough relaxing time, initial state preparation by applying $\pi/2$ pulse to spin $F_3$ (and $\pi$ pulses to spin $F_{1,2}$ after preparation of a PPS), and measurement of the FID signal of $F_3$. Typical FID and spectrum of the three spins are presented in Appendix~\ref{apd:fid}. To distinguish clearly the FIDs against the large background noise, we average 70 times the FID signals before performing the FFT to obtain the final spectrum. We deliberately configure the system off its sweet spot in order to demonstrate the advantages of physical spectral decomposition and FRESURE.

\begin{figure}[t]
	\centering
	\includegraphics[width=3in]{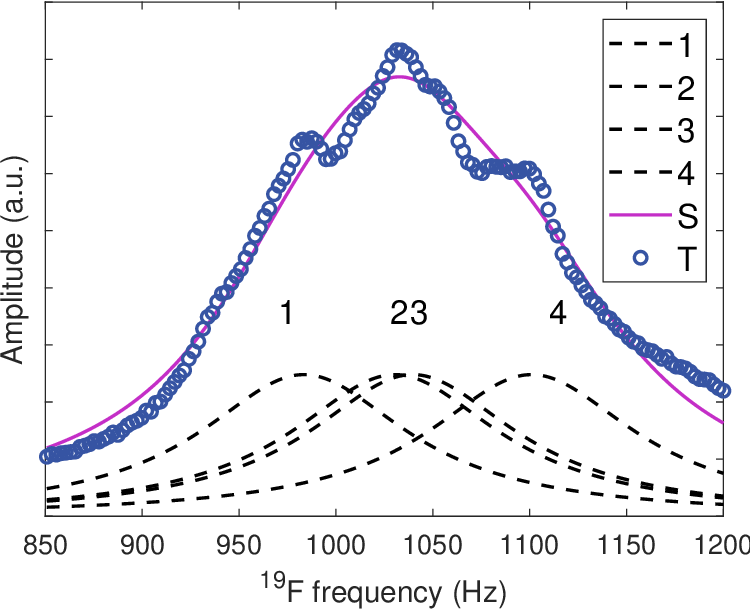}
	\caption{Fitting results of the thermal spectrum (blue circles) with four Lorentzian peaks (black dashed lines) by assuming the same width and height. The sum (purple solid line) shows only a single fat peak.}
	\label{fig:samewh}
\end{figure}
	
Experimental results for the thermal initial state are shown in Fig.~\ref{fig:samewh}, where $\tilde s_T(\omega) = FFT(s_T(t))$. We roughly observe three small peaks on a high plateau. According to conventional method, we decompose numerically the thermal initial state results into four Lorentzian peaks with the constraints of the same height and width, by inputting the system-parameter-simulated $\omega_i$ as initial value and further requiring the fitting parameters within the range $[\omega_i -10, \omega_i +10]$ Hz. The fitting results are shown in Fig.~\ref{fig:samewh} and in Table~\ref{tab:pps} (see Appendix~\ref{apd:dmfit} for other fitting results with program ``dmfit''~\cite{DMFIT}). Not surprisingly, as shown in Fig.~\ref{fig:samewh}, the sum of the decomposed four peaks exhibits only a single fat peak, not close to the experimental results at all. Clearly, numerical fitting methods are unable to correctly decompose the thermal experimental data.

\begin{figure}[t]
	\centering
	\includegraphics[width=3.25in]{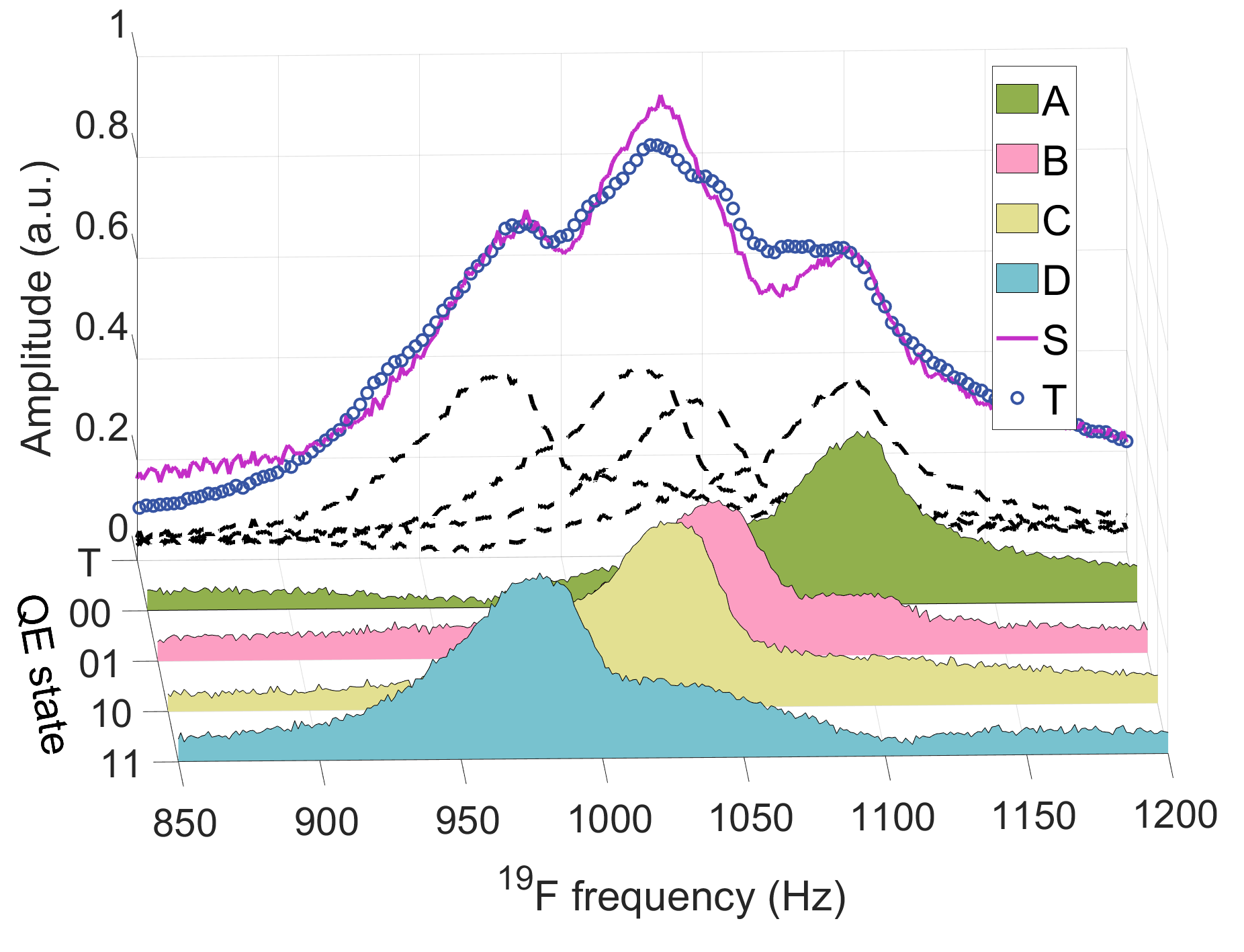}
	\caption{Experimental results of physical spectral decomposition. The lines and symbols are the same as those in Fig.~\ref{fig:numerical}. Clearly, the three peaks of the thermal state can be decomposed into four PPSs.}
	\label{fig:experiment}
\end{figure}

To demonstrate the physical spectral decomposition, we plot experimental results from the thermal initial state and four PPSs in Fig.~\ref{fig:experiment}. For each PPS, we observe a single peak, $\tilde s_{A, B, C, D}(\omega) = FFT(s_{A, B, C, D}(t))$. The peak position and width of the four PPSs are listed in Table~\ref{tab:pps}. To check the validity of physical spectral decomposition, we multiply the sum of the four PPSs' spectrum by a constant $\lambda$. An optimal $\lambda$ is searched by the least square fit, i.e., minimizing the value of $\int_{\omega_-}^{\omega_+} (\tilde s_T- \lambda \tilde s_S)^2 d\omega$ where $\omega_- = 850$ Hz, $\omega_+ = 1200$ Hz, and $\tilde s_S = \tilde s_A + \tilde s_B + \tilde s_C + \tilde s_D$. We find $\lambda$ is $1.5$ which may depend on the specific preparation process of PPSs. 

Obviously, the sum spectrum agrees well with the thermal spectrum in Fig.~\ref{fig:experiment}, indicating that the physical spectrum decomposition is valid for the $C_2F_3I$ spin system. More importantly, the spectral distance between peaks B and C is much smaller than their widths. It is impossible to discern the peaks B and C with conventional methods. However, these peaks are completely separated by introducing the extra dimension of the QE state, as shown in Fig.~\ref{fig:experiment}. Therefore, we demonstrate the FRESURE experimentally by the QE state engineering. 
	
\begin{table}[b]
	\renewcommand{\arraystretch}{1.5}
	\setlength{\tabcolsep}{4.5pt}
	\begin{center}
		\caption{Peak positions $\omega_i$ and FWHM $\Gamma_i$ ($\Gamma_i^L$) of the spectrum of PPSs, extracted from experimental results (as shown in Fig.~\ref{fig:experiment}) and the fitting results (as shown in Fig.~\ref{fig:samewh}). The units are in Hz.}
		\label{tab:pps}
		\begin{tabular}{c|c|cccc}
			\hline	\hline
			\multicolumn{2}{l|}{} & A & B & C & D \\	
			\hline
			\multirow{3}{*}{Exp.} & $\omega_i$ & 1104.7$\pm$0.6 & 1048.6$\pm$0.7 & 1024.2$\pm$0.5 & 977.8$\pm$0.7 \\
			& $\Gamma_i$ & 58.6 & 57.4 & 62.3 & 65.9 \\
			& $\Gamma_i^L$ & 57.5 & 61.6 & 60.5 & 67.8 \\
			\hline
			\multirow{2}{*}{Fit.} & $\omega_i$ & 1101.0 & 1039.6 & 1030.5 & 983.4 \\
			& $\Gamma_i$ & 65.3 & 65.3 & 65.3 & 65.3 \\
			\hline
			\hline
		\end{tabular}
	\end{center}
\end{table}
		
Compared with the numerical results presented in Fig.~\ref{fig:numerical}, the three peaks of the thermal initial state in Fig.~\ref{fig:experiment} are not distinct. Two side peaks are not symmetric with respect to the central one. The mechanism behind such an asymmetry of the side peaks is unclear and worthy exploring in the future. In addition, the experimental results for the PPSs are not exactly in Lorentzian shape because of their long and flat tail. To be more specific, we list in Table~\ref{tab:pps} the Lorentzian width $\Gamma_i^L$, which is defined as the fitting width of each PPS experimental data above 70\% of its height (to avoid their exceptional long tail), assuming a Lorentzian line shape. The validity of physical decomposition with non-Lorentzian spectral profile of the PPSs may indicate independence of the decomposition method on line-shapes.

\begin{figure}[tb]
	\centering
	\includegraphics[width=3in]{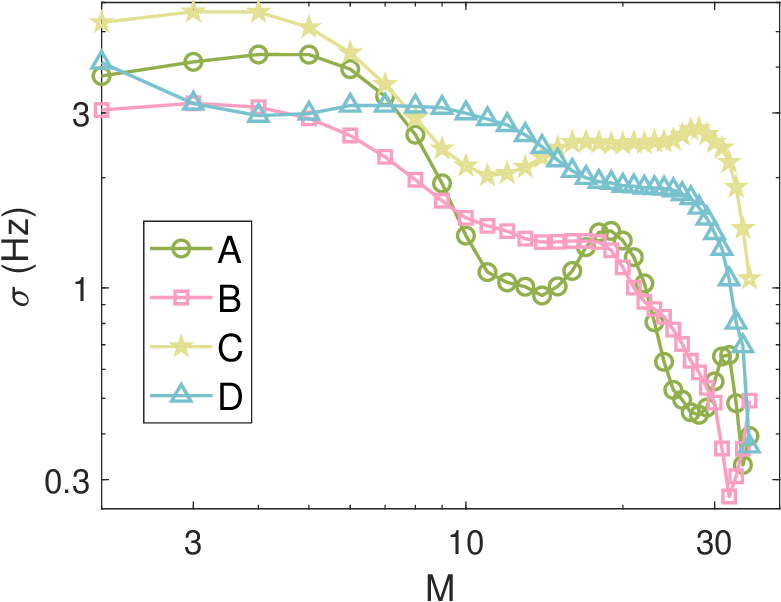}
	\caption{Dependence of Allan deviation ($\sigma$) of four PPSs on the number of averaged measurements ($M$).}
	\label{fig:allan}
\end{figure}

To explore the FRESURE limit, we calculate the Allan deviation of the peak positions of four PPSs with the Equation (4) from~\cite{Allan}.The results are presented in Fig.~\ref{fig:allan}. Clearly, as the measurements are repeated many times, the Allan deviation decreases rapidly and reaches its minimum which is in general below 1 Hz. Finally, the Allan deviation increases as the total measurement time becomes too long. Apparently, the FRESURE limits, 1 Hz in the worst case and 0.3 Hz ($\sim0.005\,\Gamma$) in the best case, are well below the peak width $\Gamma_i \sim 60$ Hz.  

We demonstrate the method of FRESURE in a rather small and simple nuclear spin system. With more advanced quantum control techniques and stronger bias magnetic fields, it is of great potential to explore the FRESURE method in more complicated but practical systems. A possible strategy for a complicated large system is divide-and-conquer, i.e., first treating the close and weakly coupled spins, then the remote spins with weaker coupling strength. 
	
	
\section{Conclusions and discussions}
\label{sec:conc}
To summarize, we propose a physical method to decompose a multiple-peak spectrum into several single peaks by the QE state engineering technique. Numerical simulations (under the original Hamiltonian Eq.~(\ref{eq:orginalH})) and experimental results in $C_2F_3I$ nuclear spin system (with SpinQ Triangulum) confirm the validity of the physical method in the weak coupling regime. With this method at hand, it is possible to realize spectral frequency super-resolution. The limit of the frequency super-resolution reaches $ \sim 0.005\; \Gamma$ in experiments, with $\Gamma$ the FWHM of the single peaks. Such a frequency super-resolution realized by QE engineering method may be useful to measure the spin coupling constants in noisy and low-field NMR systems. For example, the spin coupling strength $J_{13}$ and $J_{23}$ can be calculated from the accurate values of four PPSs' peak positions in Fig.~\ref{fig:experiment}.
	
Although the physical spectral decomposition theory assumes a weak coupling between the system and its QE, our numerical simulations and experimental results indicate that this theory is still valid with the original Hamiltonian Eq.~(\ref{eq:orginalH}) at the not-so-small value $|J_{ij}/(\delta_i-\delta_j)| \sim 0.1$. Additionally, it is worth exploring in the future the asymmetry and the long tail properties of the spectra of four PPSs, though the $\Gamma_i$ and $\Gamma_i^L$ in Table~\ref{tab:pps} are quite close.
 
Compared with spectral line narrowing, the line width of a single peak in FRESURE keeps untouched. As shown in Table~\ref{tab:pps}, the line width in experiment after FRESURE is still around 60 Hz and much larger than the distance between peaks B and C ($\sim 24$ Hz). While in line narrowing methods, such as dynamical decoupling (including spin echo), dynamic nuclear polarization, or nuclear spin state narrowing, the line width of each single peak is strongly reduced, so that the peak distance eventually becomes larger than the reduced line width. Additionally, the limit of line width by line narrowing,such as dynamical decoupling or environmental state narrowing, is in the order of $T_1^{-1}$ with $T_1$ the longitudinal spin relaxation time~\cite{opticallinenarrowing, Dourenarrow}. For the $C_2 F_3 I$ nuclear spin system, $T_1\sim 10$~s and the narrowest line width is about 0.1 Hz~\cite{Dujiangfeng2016, dujiangfeng2021}. Obviously, the resolution with FRESURE ($\sim 0.3$ Hz) is close to the limit of line narrowing techniques, but the experimental condition is significantly relaxed, e.g., lower bias field and noisier stray field fluctuation.
	
	
\begin{acknowledgments}
	The work is supported by National Natural Science Foundation of China (NSFC) under Grant No. 12274331,  Innovation Program for Quantum Science and Technology under Grant No. 2021ZD0302100, and the NSAF under Grant No. U1930201. The numerical calculations in the paper have been partially done on the supercomputing system in the Supercomputing Center of Wuhan University.
\end{acknowledgments}

\appendix

\section{Mathematical ``diffraction'' and its limit in NMR spectrum}
\label{apd:diffraction}	

In optics, a single point light source spreads on the image plane into an Airy disc described by point spread function (PSF)~\cite{Jiang2019}, after passing through a finite size lens which causes point spreading due to the diffraction. For example, X-ray diffraction and NMRd are such a spatial diffraction~\cite{NMRd}, which is different from our FRESURE in spectrum. Whether two point light sources are spatially resolvable depends on the diffraction limit (effective size of the PSF) described by Abbe-Rayleigh criterion~\cite{doi:10.1142/5833}. Although no optical diffraction occurs in NMR spectrum, there are Fourier transform and spectral line broadening, which is nothing but ``spectral PSF''. In this mathematical sense, it is reasonable to draw an analogy between the spectral line broadening and optical diffraction. The spectral resolution limit of a single peak position is determined mathematically by the Fourier transform limit. For two or multiple peaks, similar to the optical diffraction limit, the spectral resolution limit in NMR depends on the difference between peak distance  and the effective PSF size (line width), approximately the transversal spin relaxation rate $1/T_2^*$~\cite{slichter1996principles}. In the case of larger peak distance than the PSF size, the peak-position precision may approach the Fourier transform limit. While in the rest cases, it is approximately the PSF size.

In our experiments, we consider four spectral peaks. Two central ones of them lie closely and within the line width (under ``diffraction limit”). It is challenging to discern them without the knowledge of PSF. Unfortunately, the PSF is unknown in our NMR spin system (neither Lorentzian or Gaussian function). Consequently, the widely-used multiple peak fitting program ``dmfit” fails, as shown in 
Fig.~\ref{fig:dmfitexp34} in Appendix~\ref{apd:dmfit}. In a contrast, the FRESURE method, intrigued by the idea of super-resolution in PALM or STORM~\cite{HUSZKA20197, Hell_2015, rust2006sub,Bo2008, palmBetzig}, is able to beat the ``spectral diffraction limit” and achieves approximately the Fourier transform limit.

\section{FID and FFT signal in experiments}
\label{apd:fid}

\begin{figure}
	\centering
	\includegraphics[width=3.25in]{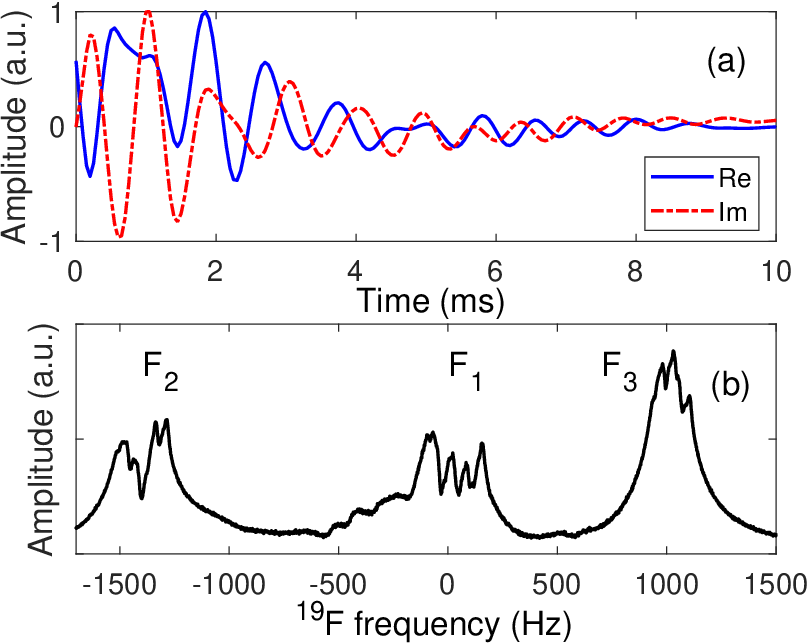}
	\caption{(a) Typical FID signal (averaged over 100 times) and (b) its spectrum (amplitude). Real (imaginary) part of the FID is plotted as blue-solid (red-dotted) line. Three main peaks, from left to right, denote the spectra of $F_2$, $F_1$, and $F_3$, respectively. Each main peak in principle consists of four subpeaks. In experiment, we adjust the second subpeak of $F_1$ at $0$ Hz.}
	\label{fig:FID}
\end{figure}

The FID signal in Fig.~\ref{fig:FID}(a) illustrates the time dependence of total magnetization of three fluorine nuclear spins, including the real and imaginary part (the total acquisition time is 0.8 s). Obviously, such signals deviates from an ideal FID of a single spin, exponential decay, due to the complex interaction between three fluorine spins.

From Fig.~\ref{fig:FID}(b), the FFT of the FID signal, the chemical shifts $\delta_{1, 2,3}$ and the spin coupling strengths $J_{12}, J_{13}$, and $J_{23}$  are extracted and estimated from the peak positions. Typical peak width $\Gamma$ (FWHM) is around $60$ Hz from the spectrum of $F_1$.

\section{Spectral decomposition of numerical and experimental results with ``dmfit'' program}
\label{apd:dmfit}

\begin{figure}[tb]
	\centering
	\includegraphics[width=3.25in]{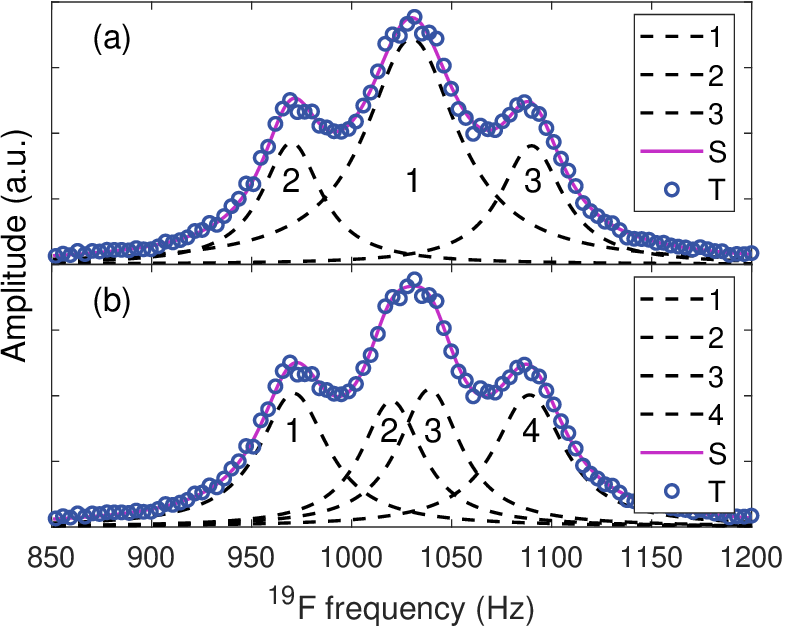}
	\caption{Spectral decomposition of the thermal initial state (blue circles), obtained by numerical simulations with the Hamiltonian Eq.~(\ref{eq:orginalH}) in Section~\ref{sec:derive}, with the ``dmfit" program~\cite{DMFIT} and assuming (a) three Lorentzian peaks (denoted as black dashed lines 1, 2, and 3) and (b) four Lorentzian peaks (denoted as black dashed lines 1, 2, 3 and 4). The fitted results (purple solid line) agree well with the simulation data.}
	\label{fig:dmfitnum34}
\end{figure}

The thermal spectrum may also be decomposed numerically with the program ``dmfit''~\cite{DMFIT}, by assuming three or four Lorentzian peaks and carefully choosing initial guesses of parameters (especially peaks' position, height and width). Both three-peak and four-peak options eventually generate pretty good fitting. In NMR, the central, high and fat peak of the three-peak fitting result is considered as two nearly degenerate peaks since the central peak area is about twice the left or right.

In Fig.~\ref{fig:dmfitnum34}, we employ the program ``dmfit'' to carry out the spectral decomposition of the thermal initial state from numerical results~\cite{DMFIT}. By assuming three peaks with Lorentzian lineshape, it is easy to obtain the fitting results shown in Fig.~\ref{fig:dmfitnum34}(a). With the knowledge of the $C_2F_3I$,  we also carry out four peaks fitting, shown in Fig.~\ref{fig:dmfitnum34}(b), by carefully adjusting each peak's position and width in ``dmfit'' program. Obviously, ``dmfit" generates the almost same result as numerical simulations by four-peak fitting. Nearly perfect fitting with three- and four-peak implies that it is not flawless with numerical fitting.

\begin{table}[b]
	\renewcommand{\arraystretch}{1.5}
	\setlength{\tabcolsep}{5pt}
	\begin{center}
		\caption{Peak positions $\omega_i$ and FWHM $\Gamma_i$ of the experimental spectrum fitting of ``dmfit", extracted from Fig.~\ref{fig:dmfitexp34}. The units are in Hz.}
		\label{tab:dmfitexp}
		\begin{tabular}{c|c|cccc}
			\hline	\hline
			\multicolumn{2}{l|}{} & 1 & 2 & 3 & 4 \\	
			\hline
			\multirow{2}{*}{Three-peak fitting} & $\omega_i$ & 977.0 & 1032.9 & 1077.2 & none \\
			& $\Gamma_i$ & 79.7 & 47.5 & 177.9 & none \\
			\hline
			\multirow{2}{*}{Four-peak fitting} & $\omega_i$ & 976.0 & 1038.3 & 1098.7 & 1147.0 \\
			& $\Gamma_i$ & 89.2 & 74.9 & 46.9 & 262.8 \\
			\hline
			\hline
		\end{tabular}
	\end{center}
\end{table}

\begin{figure}[tb]
	\centering
	\includegraphics[width=3.25in]{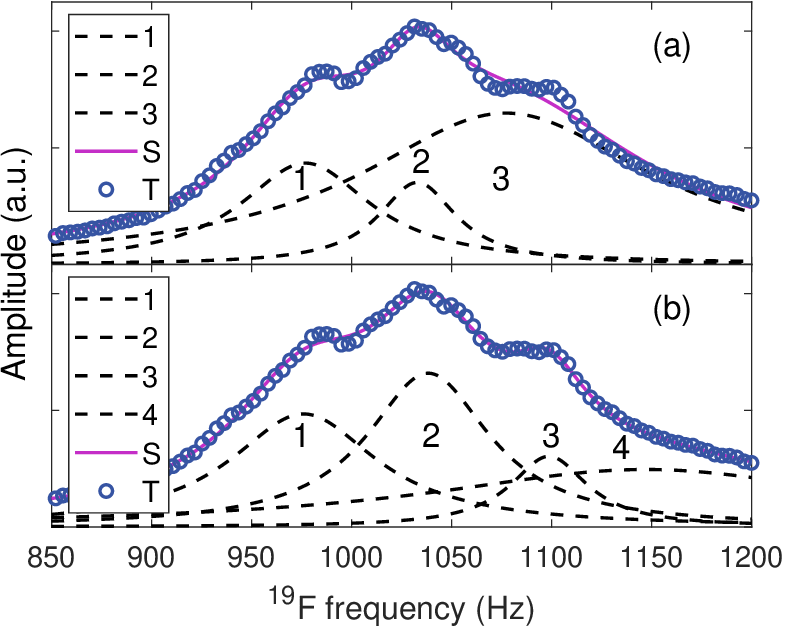}
	\caption{(a) Three- and (b) four-peak spectral decomposition of the thermal initial state in experiments with ``dmfit" program~\cite{DMFIT}.}
	\label{fig:dmfitexp34}
\end{figure}

We also use the ``dmfit'' program to fit experimental data of the spectrum of thermal initial state.  For three-peak fitting, the parameters we initially set for peak 1, 2 and 3 in Fig.~\ref{fig:dmfitexp34}(a) are 20, 10, 20 for amplitudes, 970, 1030, 1070 for positions, and 90, 40, 180 for widths, respectively. For four-peak fitting, the parameters we initially set for peak 1, 2, 3 and 4 in Fig.~\ref{fig:dmfitexp34}(b) are 20, 20, 20, 20 for amplitudes, 970, 1035, 1090, 1100 for positions, and 90, 70, 40, 200 for widths,  respectively. We tried other initial parameters but the above values generates the best fitting. Compared to the physical spectral decomposition with quantum environment engineering in the main text, both three- and four-peak ``dmfit'' fitting generate physically incorrect spectral decomposition.

\section{ Valid range of weak coupling approximation}
\label{apd:range}

\begin{figure}[b]
	\centering
	\includegraphics[width=3.25in]{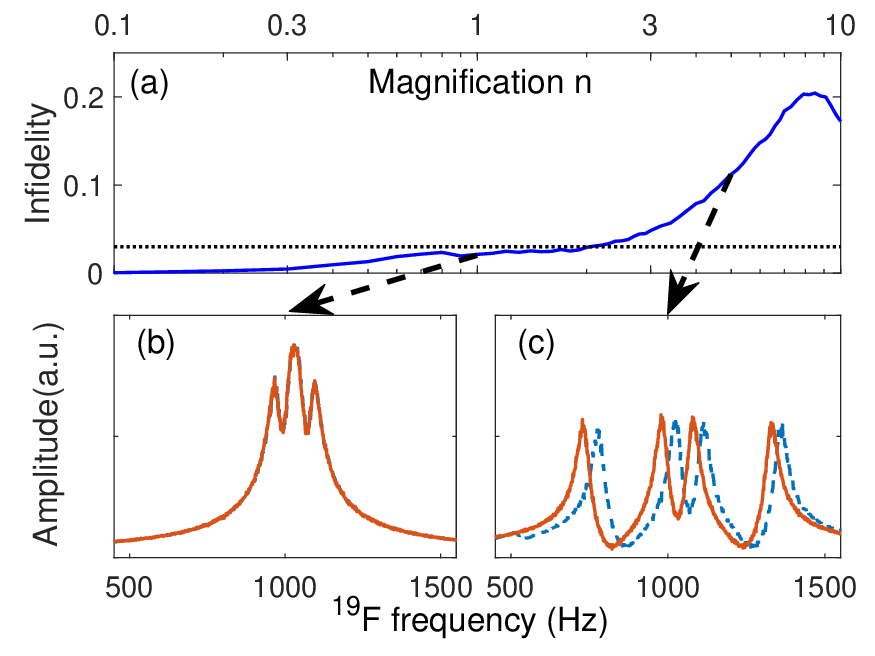}
	\caption{(a) Dependence of the spectral infidelity on the coupling strength magnification. The horizontal dotted line indicates the threshold of spectral infidelity. Typical spectra in (b) the weak coupling regime ($n = 1$) and (c)  the strong coupling regime ($n = 5$). The red solid and blue dashed lines in panels (b) and (c) denote the numerical spectral results from Eq.~(\ref{eq:scalarH}) and Eq.~(\ref{eq:orginalH}), respectively.  The weak coupling approximation is valid ($n \lesssim 2$) if the spectral infidelity is below its threshold.}
	\label{fig:range}
\end{figure}

To investigate the valid range of the weak coupling approximation, we introduce the coupling strength magnification $n$, which magnifies the coupling strength by a factor $n$, i.e., $J_{ij} \rightarrow n\times J_{ij}$. By carrying out numerical simulations with the Hamiltonians from Eqs.~(\ref{eq:orginalH}) and (\ref{eq:scalarH}), with the magnified coupling constants $n \times J_{ij}$,  we follow the same procedure as in Fig.~\ref{fig:numerical} to calculate these two spectra, $\tilde s_1(\omega)$ and $\tilde s_2(\omega)$, respectively. We then define a spectral  infidelity

\begin{equation}
\frac{\Delta S} {S} = \frac {\int |\tilde s_1-\tilde s_2| d\omega} {\int (\tilde s_1 + \tilde s_2) d\omega} \, ,
\end{equation}

which characterizes the difference between the two spectra. The infidelity parameter is zero if these two spectra are the same, and 1 if they are completely different (no overlap between them). As shown clearly in Fig.~\ref{fig:range}, the valid range of the weak coupling approximation is approximately $n \lesssim 2$.



%

\end{document}